\documentclass[graybox]{svmult}
\usepackage{comment, array, amsmath}
\usepackage{color}
\usepackage[colorinlistoftodos]{todonotes}
\usepackage[hidelinks]{hyperref}
\usepackage{amssymb, amscd, mathptmx, helvet, courier, graphicx, makeidx, multicol, footmisc, color, framed,
enumerate}

\spdefaulttheorem{project}{Research Project}{\bf}{}
\newcommand{\resproject}[1]{\begin{svgraybox} \begin{project} #1 \end{project} \end{svgraybox}}

\spdefaulttheorem{cproblem}{Challenge Problem}{\bf}{}
\newcommand{\chproblem}[1]{\begin{cproblem} #1 \end{cproblem}}

\spnewtheorem*{prerequisites}{Suggested prerequisites}{\bf}{\small\em}

\begin{document}

\title*{Phylogenetic Networks}
\author{Elizabeth Gross, Colby Long, and Joseph Rusinko}

\institute{E. Gross \at University of Hawai`i at M\={a}noa,
2565 McCarthy Mall, Honolulu, HI 96822, \email{egross@hawaii.edu} \and C. Long \at Mathematical Biosciences Institute, 1735 Neil Ave, Columbus, OH 43210,
 \email{long.1579@mbi.osu.edu} \and J. Rusinko \at Hobart and William Smith Colleges, 300 Pulteney Street, Geneva, NY 14456, \email{rusinko@hws.edu}}

\maketitle


\abstract{
Phylogenetics is the study of the evolutionary relationships between organisms. 
One of the main challenges in the field is to 
take biological data for a group of organisms
and to infer an evolutionary tree, 
a graph that represents these relationships. 
Developing practical and efficient methods for inferring phylogenetic trees has lead to a number of interesting mathematical questions across a variety of fields. However, due to hybridization and gene flow, a phylogenetic network may be a better representation of the evolutionary history of 
some groups of organisms. 
In this chapter, we introduce some of the basic
concepts in phylogenetics, and present 
related research projects on phylogenetic networks that touch on areas of graph theory and abstract algebra.
In the first section, we describe several open research questions related to the combinatorics of  phylogenetic networks. In the second, we
we describe problems related to understanding phylogenetic statistical models as algebraic varieties. 
These problems fit broadly
in the realm of algebra, but could be more 
accurately classified as problems in \emph{algebraic statistics} or \emph{applied algebraic geometry}.
}

\begin{prerequisites}
An introductory course in graph theory or discrete mathematics for the research projects
in Section \ref{sec: Combinatorics of Phylogenetic Networks}. 
For the projects in Section \ref{sec: Algebra of Phylogenetic Networks}, an introductory course in abstract algebra would also be helpful.
\end{prerequisites}

\section{Introduction.}
\label{sec-intro}

The field of phylogenetics is concerned with uncovering the evolutionary relationships between species.
Even before Darwin proposed evolution through variation and natural selection, people used family trees to show how individuals were related to one another.
Since Darwin's theory implies that all species alive today are descended from a common ancestor, the relationships among any group of individuals, even those from different species, can similarly be displayed on a phylogenetic tree.
Thus, the goal of phylogenetics is to use biological data
for a collection of individuals or species, and to infer a tree that describes how they are related.
In modern phylogenetics, the biological data that we consider is 
most often the aligned DNA sequences for the species under consideration.
Understanding how species have evolved has important applications in
evolutionary biology, species conservation, and epidemiology \cite{Semple2003}.

Perhaps unsurprisingly, there is a rich interplay between phylogenetics and mathematics.
A tree can be viewed as a certain type of graph,
and graph theory is an entire field of mathematics dedicated to understanding the structure and properties of graphs. 
Similarly, DNA mutation is a random process, and understanding random processes falls in the domain of probability and statistics.
As such, there are many mathematical tools that have been
developed for doing phylogenetic inference. Often, developing a new tool or trying to answer a novel question in phylogenetics requires solving 
some previously unsolved mathematical problem.
It is also common for a phylogenetic problem to suggest a mathematical problem that is interesting in its own right.

The outline above, where every set of species is related by a phylogenetic tree, is a simplified description of the evolutionary process. 
Rarely does the evolutionary history for a set of species neatly conform to this story.
Instead, species hybridize and swap genes. Moreover, genes within individuals have their own unique evolutionary histories that can differ from that of the
individuals in which they reside \cite{Maddison1997, syvanen1994}. 
The result is that in many cases, a tree is simply insufficient to represent the evolutionary process.
Recognizing this, many researchers have argued that networks can be a more appropriate way to represent evolution.
While using networks might be more realistic from a biological standpoint, there are many complexities and new mathematical questions that must be solved in order to infer phylogenetic networks. 
In particular, understanding inference for networks requires proving results for networks analogous to those known for trees. The projects that we present in this chapter are examples of some of the new lines of inquiry inspired by using networks in phylogenetics.

The first category of problems that we describe
concern the combinatorics of phylogenetic networks.
Inferring phylogenies for large sets of species
can often be computationally intensive regardless
of the method chosen. 
One approach for dealing with this in the tree
setting is to consider small subsets of species 
one at a time. Once phylogenetic trees have been 
built for each subset, the small trees are then
assembled to construct the tree for the entire
set of species. 
The details of actually doing this can of course become quite complicated. Thus, different heuristics and algorithms have been proposed, and understanding their performance and properties leads to a number of interesting questions about the combinatorics of trees. As a first example, one might consider if it
is even possible to uniquely determine the species tree for a set of species only from knowledge of how each subset of a certain size is related. 
Even if this is possible, one then might like to know
how to resolve contradictions between subtrees
if there is error in the inference process.
Adopting a similar strategy for inferring phylogenetic
networks from subnetworks leads to a host of 
similar combinatorial questions about networks.
In Section \ref{sec: Combinatorics of Phylogenetic Networks}, we 
will explore the structure of phylogenetic networks in greater depth and formulate some of these questions more precisely for potential research projects.

The second class of problems we discuss concerns the surprising connections between abstract algebra and phylogenetics.
One of the ways that researchers have sought
to infer phylogenies is by building models
of DNA sequence evolution on phylogenetic trees.
Once the tree parameter is chosen, the numerical parameters of the model control the rates and types of mutations that can occur as evolution proceeds along the tree.
Once all the parameters for the model are specified,
the result is a probability distribution on DNA site-patterns. 
That is, the model predicts the frequency with which different DNA site-patterns will appear in the aligned DNA sequences of a set of species. For example, the model might predict that at the same DNA locus for three species, there is a 5\% chance that the DNA nucleotide {\tt{A}} is at that locus in each species. 
Another way to write this is to write that for this choice of parameters, $p_{AAA} = .05$.
Algebra enters the picture when we start to consider the 
algebraic relationship between the predicted sight pattern frequencies. For example, we might find that for a particular 
model on a tree $T$, no matter how we choose the numerical
parameters the probability of observing {\tt{ACC}} under the model is always the same as the probability of observing {\tt{GTT}}. 
We can express this via the polynomial relationship $p_{ACC} - p_{GTT} = 0$, and polynomials that always evaluate to zero on the model we call \emph{phylogenetic invariants} \cite{Cavender} for the model on $T$.

The set of all phylogenetic invariants for a model is an algebraic object called an \emph{ideal}.
By studying the ideals and invariants associated to phylogenetic models, researchers have been able to prove various properties of the models, such as their dimension and whether or not they are identifiable, as well as to develop new methods for phylogenetic inference 
(see e.g., \cite{Allman,Casanellas2006,Chang1996, Rusinko2012}).
As with some of the combinatorial questions above, there are a number of papers studying these questions in the case of trees, but few in the case of networks. 
In Section 
\ref{sec: Algebra of Phylogenetic Networks},
we show how to associate invariants and ideals to phylogenetic networks and describe several related research projects. 
While there are fascinating connections between these algebraic objects and statistical models of DNA sequence evolution, our presentation distills some of the background material and emphasizes
the algebra. There is also a computational algebra component to some of these projects
and we provide example computations with Macaulay2 \cite{M2} code.

\section{Combinatorics of Phylogenetic Networks}
\label{sec: Combinatorics of Phylogenetic Networks}

In this section, we give the background necessary to work on the research and challenge questions related to the structure of phylogenetic networks. 
We begin by introducing some of the concepts 
from graph theory necessary to formally define a phylogenetic tree and a phylogenetic network. 
We then discuss some ways to encode
trees and networks and common operations that 
we can perform on them.
Much of the terminology around trees and graphs is standard in graph theory, and so we have
omitted some of the basic definitions that can 
be found in the first chapter of any text on the subject. 
An affordable and helpful source for more information and standard graph theoretic results would be 
\cite{chartrand2013first}. 
 The terms that are specific
to phylogenetic trees and networks we 
have adapted largely from
\cite{Francis2016, Semple2016}.
The textbook \cite{huson2010phylogenetic} provides a thorough introduction to phylogenetic networks, though the specific terminology being used in the research literature is still evolving.  A broader introduction to phylogenetics from a mathematical perspective can be found in \cite{Steel2016Phylogeny}.

\subsection{Graphs and Trees}

The outcome of a phylogenetic analysis is 
typically a phylogenetic tree, a graph that describes the ancestry for a set of taxa. 
As an example,
an interactive phylogenetic tree relating
hundreds of different species can be accessed
at 

\begin{center}
    {\tt{https://itol.embl.de/itol.cgi}}.    
\end{center}

In mathematical terms, a \emph{tree} is a connected graph with no cycles. 
We refer to the degree one vertices of a tree
as the \emph{leaves} of the tree. 
The leaves correspond to the extant species for
which we have data in a phylogenetic analysis
and so we label these vertices by some label set. In theoretical applications, the label set for an $n$-leaf tree is often just the set 
$[n] := \{1, \ldots, n\}$, and we call 
such a tree an \emph{$n$-leaf phylogenetic tree}.
Note that we consider two $n$-leaf phylogenetic trees to be distinct even if they differ only
by the labeling of the leaves.
In technical terms, two $n$-leaf trees are the same if and only if there is a graph isomorphism between them that also preserves the leaf-labeling. 

We often distinguish one special vertex of an
$n$-leaf phylogenetic tree which we call 
the \emph{root}. 
If the root is specified, then the tree can be regarded as a directed graph, where all edges point away from
the root. The root corresponds to the common ancestor of all of the species of the tree, hence, the directed edges can be thought of as indicating the direction of time.
We also often restrict the set of trees we consider to
those that are \emph{binary}. 
A binary tree is one in which every vertex other than the root has degree one or degree three. If the root is specified for a binary tree, then it will have degree two. We use these rooted binary phylogenetic trees as a model of evolution. 
The degree three internal vertices
correspond to speciation events, where there is 
one species at the time just prior to the vertex, 
and two species that emerge from the vertex.

Depending on the application, it is common in phylogenetics to consider both rooted and unrooted trees. As such, we can think about \emph{rooting} a tree, 
where we place a degree two vertex along an edge and direct the edges away from this vertex (so that there is a directed path from the root to every vertex in the graph). Or, we can
think about \emph{unrooting} a tree, where we suppress the degree two vertex (see the trees in Figure \ref{fig: TreeExample}). As an example, there is only one 3-leaf binary phylogenetic tree, however, there are three different rooted 3-leaf binary phylogenetic trees that can be obtained by rooting along the three different edges of the unrooted tree.

\begin{example}
\label{ex: TreeExample}

Figure \ref{fig: TreeExample} shows
a rooted 4-leaf binary phylogenetic tree and
the tree obtained by unrooting this tree.
Notice that the edges of the rooted tree are directed, but that this is unnecessary since
the root determines the direction of each edge.
Also observe that if we root the unrooted tree along the edge labeled by $1$, we obtain the rooted tree at left. Finally, notice that 
swapping the labels $1$ and $3$ in the rooted tree produces a distinct rooted 4-leaf binary phylogenetic tree, 
whereas for the unrooted tree, swapping these labels leaves the tree unchanged. 

\begin{figure}[h]
\begin{center}
\includegraphics[width=3in]{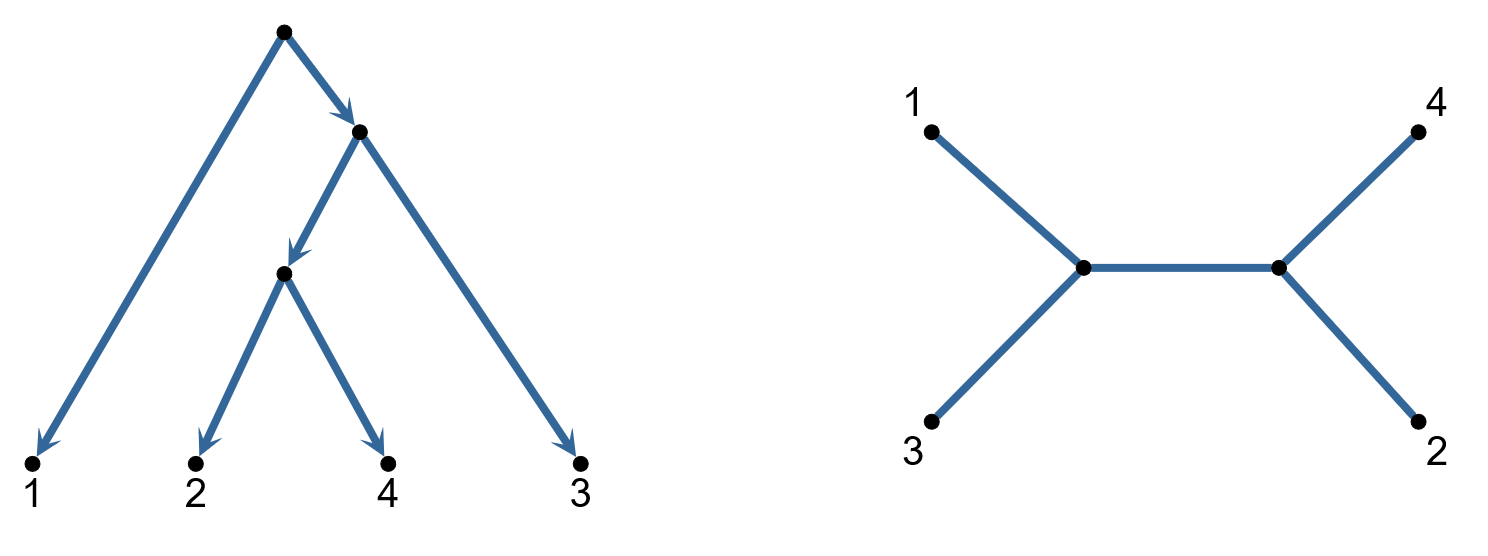}
\caption{
    \label{fig: TreeExample} 
    A rooted 4-leaf binary phylogenetic tree and
    the tree obtained by unrooting this tree. 
} 
\end{center}
\end{figure}

\end{example}

\begin{exercise}
How many edges are there in an
$n$-leaf rooted binary phylogenetic 
tree?
\end{exercise}

\begin{exercise}
Prove that there exists a unique path between any pair of vertices in a tree.
\end{exercise}

\begin{exercise}
Prove that the number of 
rooted binary phylogenetic $n$-trees
is (2n - 3)!! Here the symbol $!!$ does not mean the factorial of the factorial, but rather multiplying by numbers decreasing by two.  For example, $7!!=7\times5\times3\times 1=105$ and $10!!=10\times 8 \times 6 \times 4 \times 2=3840$.
\end{exercise}

Each edge of an unrooted phylogenetic tree subdivides the collection of leaves into a pair of disjoint sets.  This pair is called a split. For example the unrooted tree in Figure \ref{fig: TreeExample} displays the splits $S=\{1|234,2|134,3|124,4|123, 13|24\}$.

\begin{exercise}
Draw an unrooted tree with the set of splits  $$S=\{1|23456,2|13456,3|12456,4|12356,5|12346,6|12345,  13|2456,135|246,46|1235\}.$$
\end{exercise}

\chproblem{Prove that two unrooted phylogenetic trees are isomorphic if and only if they display the same set of splits.}

\subsection{Phylogenetic Networks}
\label{sec: Phylogenetic Networks}

As mentioned in the introduction, a tree might not 
always be sufficient to describe the history of a set
of species. For example, consider the graphs depicted in Figure \ref{fig: 3SemidirectedNetworks}.
Notice that there are vertices in these graphs 
with in-degree two and out-degree one. 
There are a few ways that we might interpret these
\emph{reticulation events}. 
It could be that two distinct species entered the vertex, and only one, their hybrid, emerged.
Or, it might be that one of the edges directed into the degree two vertices represents a gene flow event where species remain distinct but exchange a small
amount of genetic material.
If we undirect all of the edges of either of these
graphs, the result is clearly not a tree since the
resulting undirected graph contains a cycle. 
In fact, this is a phylogenetic network.
A more thorough introduction
to phylogenetic networks than we offer 
here can be found in 
\cite{huson2010phylogenetic,Nakhleh2011}.
The website ``Who's who in phylogenetic
networks" \cite{Agarwal2016Whoiswho} is also
an excellent resource for discovering articles
and authors in the field.

\begin{definition}
\label{def: network}
A phylogenetic network  
$N$
 on a set of leaves $[n]$ is a rooted acyclic directed graph with no edges in parallel (i.e. no multiple edges) and satisfying the following properties:
\begin{enumerate}[(i)]
\item The root has out-degree two.
\item The only vertices with out-degree zero are the leaves $[n]$ and each of these have in-degree one.
\item All other vertices either have in-degree one and out-degree two, or in-degree two and out-degree one.
\end{enumerate}
\end{definition}

\begin{figure}[h]
\begin{center}
\includegraphics[width=4in]{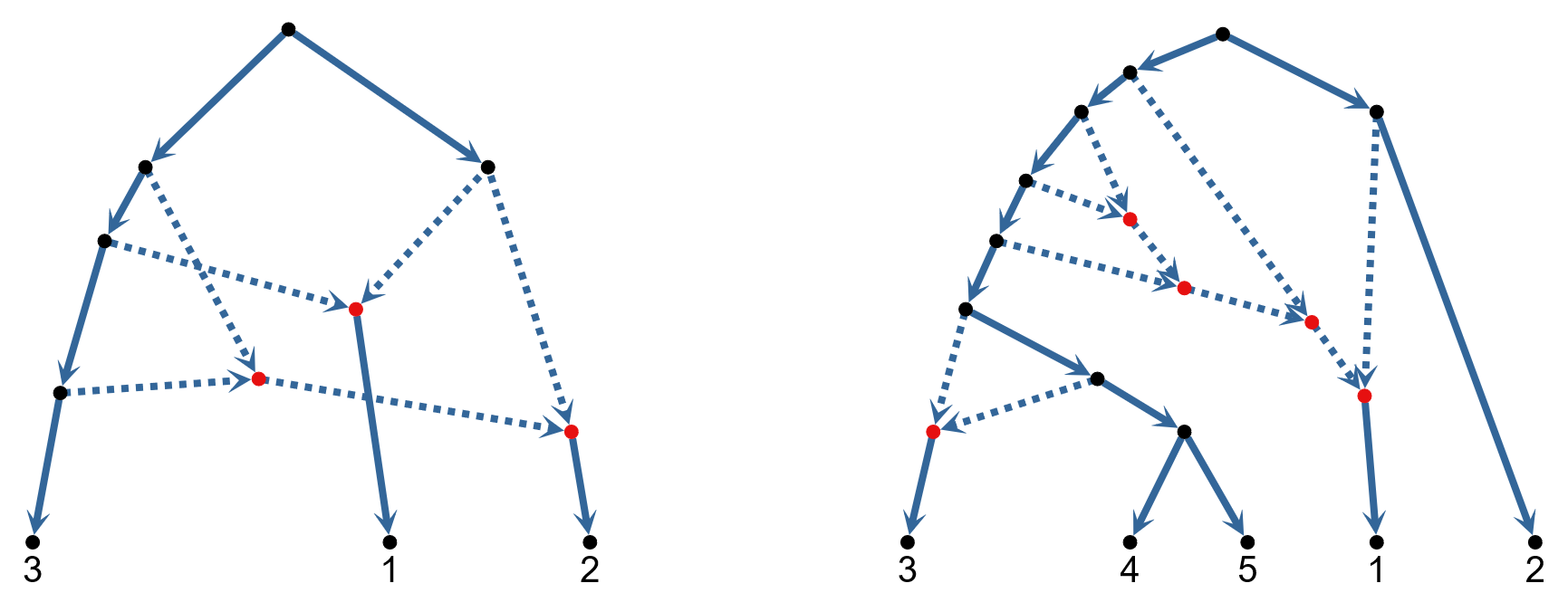}
\caption{
\label{fig: TwoPhylogeneticNetworks} 
Two rooted binary phylogenetic networks.
} 
\end{center}
\end{figure}

In the preceding definition, the 
term acyclic refers to the fact that the
network should contain no \emph{directed} cycles.
The vertices of in-degree two are called the  \emph{reticulation vertices} of the network
since they correspond to reticulation events.
Likewise, the edges that are directed into
reticulation vertices are called 
\emph{reticulation edges}.
Observe that the set of rooted binary phylogenetic trees are subset of the set of phylogenetic networks.

\begin{exercise}
For what $m \in \mathbb{N}$ is it possible
to draw a rooted 3-leaf phylogenetic network with exactly $m$ edges?
\end{exercise}

\begin{exercise}
Show that there are an infinite number of rooted $n$-leaf phylogenetic networks.
\end{exercise}

The ability of phylogenetic networks
to describe more complicated evolutionary histories comes at a cost in that networks can be much more difficult to analyze. Since there are infinitely many phylogenetic networks versus only finitely many phylogenetic trees, selecting the best network to describe a set of species is particularly challenging.
Because there are so many networks,
it is often desirable to consider
only certain subclasses of phylogenetic
networks depending on the particular
application. 
One way to restrict the class of networks is
by considering only networks with a 
certain number of reticulations or 
those of a certain 
\emph{level}.
The concept of the level of a network, introduced in \cite{Jansson2006Inferring}, 
relies on the definition of
a \emph{biconnected component}
of a graph.

\begin{definition}
\label{def: biconnected}
A graph $G$ is \emph{biconnected} 
(or 2-connected) if for every vertex
$v \in V(G)$, $G - \{v\}$ is a 
connected graph.
The \emph{biconnected components} of a graph
are the maximal biconnected subgraphs.
\end{definition}

\begin{definition}
    The \emph{reticulation number} of a phylogenetic network is the total number of reticulation vertices of the network.
    The \emph{level} of a rooted phylogenetic network is the maximum number of reticulation vertices in a biconnected
    component (considered as an undirected
    graph) of the network.
\end{definition}

\begin{exercise} 
What is the reticulation number of the two phylogenetic networks pictured in Figure \ref{fig: 3SemidirectedNetworks}?  What are the biconnected components of each network? What are the levels of the two networks?
\end{exercise}

\begin{exercise}
\label{exer: remove reticulation edges}
Suppose that you remove one reticulation
edge from each pair of edges directed
into a reticulation vertex in a 
phylogenetic network. Show that if 
you undirect the remaining edges the result is a connected, acyclic graph.
\end{exercise}

\begin{exercise} 
\label{exer: how many rooted level-one networks?}
How many 3-leaf rooted phylogenetic networks with
a single reticulation vertex are there?
\end{exercise}

\subsection{Semi-directed Networks}

Whether we work with rooted or unrooted phylogenetic
trees or networks
depends upon the particular application. 
As an example, for some
statistical models of DNA sequence evolution,
the models for two distinct rooted trees will be the same if the trees are the same when unrooted. 
Thus, when working with such models, there is no basis for selecting one location of the root over any other, and so we only concern ourselves with 
unrooted trees.

Rooting a tree is one way of assigning a direction to each of its edges.  When constructing evolutionary models associated to phylogenetic networks it can occur that the direction of some edges can be distinguished by the model, but that the directions associated to other edges can not. 
Thus it makes sense to consider the class of unrooted networks in which some of the edges are directed which are known as \emph{semi-directed} networks.

For certain algebraic models of evolution, the models will not necessarily be the same
if the unrooted phylogenetic network parameters are the same. However, they will be if the underlying \emph{semi-directed  topology} of the networks is the same.

\begin{definition}
The \emph{semi-directed topology} of a rooted 
phylogenetic network is the semi-directed network obtained by unrooting the network and undirecting all non-reticulation
edges.
\end{definition}

Because of the increasing importance of networks in phylogenetics, several authors have investigated the
combinatorics of both rooted and unrooted phylogenetic networks (e.g., \cite{Francis2016, Pardi2015, VanIersel2014}). 
The semi-directed topology has
recently appeared in some applications \cite{Gross2017distinguishing, Solis2016},
but the combinatorics of these networks
have received comparatively little attention. 

Of course, the semi-directed networks 
that we are interested in are those that 
actually correspond to the semi-directed 
topology of a rooted phylogenetic network, which 
we call \emph{phylogenetic semi-directed networks}.
An edge in a phylogenetic semi-directed network
is a \emph{valid root location} if the network
can be rooted along this edge and orientations
chosen for the remaining undirected edges
to yield a rooted phylogenetic network.

\begin{example}
Figure \ref{fig: 3SemidirectedNetworks}
shows three semi-directed networks.
The 3-leaf semi-directed network is a phylogenetic
semi-directed network, which can be seen by noting
that it is the semi-directed topology of the
3-leaf network in 
Figure \ref{fig: TwoPhylogeneticNetworks}.
The 4-leaf semi-directed network is not a phylogenetic semi-directed network. Notice that there is no way to orient the edge connecting the
reticulation vertices without creating vertices
of in-degree 3 and out-degree 3, violating
the conditions of Definition \ref{def: network}.
The 6-leaf network is also a phylogenetic semi-directed
network (Exercise \ref{exer: is phylogenetic}).

\begin{figure}[h]
\begin{center}
\includegraphics[width=3in]{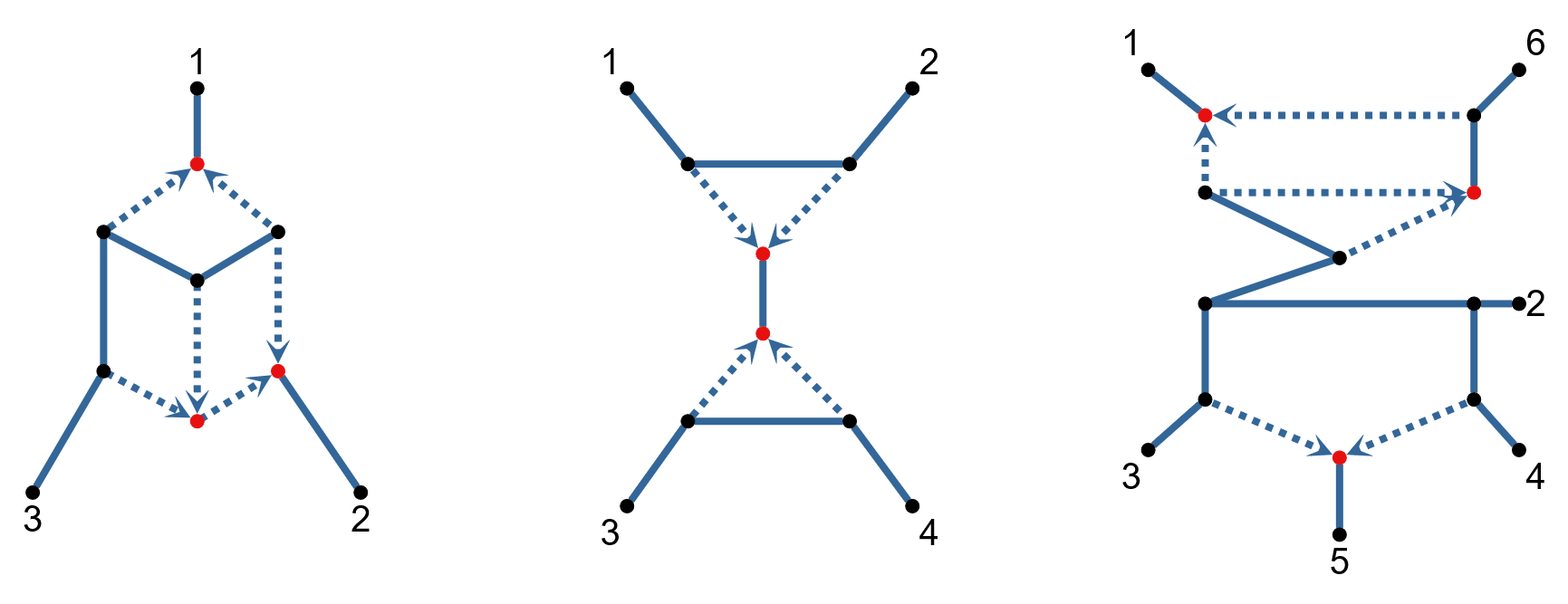}
\caption{
\label{fig: 3SemidirectedNetworks} 
A 3-leaf phylogenetic semi-directed network,
a 4-leaf semi-directed network that is
not phylogenetic, and a 
6-leaf phylogenetic semi-directed network.}
\end{center}
\end{figure}

\end{example}

\begin{exercise} 
Find all of the valid root locations
for the 3-leaf phylogenetic 
semi-directed network in Figure 
\ref{fig: 3SemidirectedNetworks}.
\end{exercise}

\begin{exercise} 
\label{exer: is phylogenetic}
Show that the 6-leaf semi-directed network in Figure \ref{fig: 3SemidirectedNetworks} 
is a phylogenetic semi-directed network.
Find all of the valid root locations.
\end{exercise}

\begin{exercise} 
Draw the semi-directed topology of the
5-leaf rooted phylogenetic network in 
Figure \ref{fig: 3SemidirectedNetworks}.
\end{exercise}

\begin{exercise} 
How many 4-leaf semi-directed networks with a single reticulation are there?
\end{exercise}


\begin{exercise}
\label{exer: not phylogenetic}
Show that there is no way to direct 
any of the existing undirected edges 
in the semi-directed network 
in Figure \ref{fig: ExerciseSemidirected}
below to obtain a phylogenetic semi-directed network.

\begin{figure}[h]
\begin{center}
\includegraphics[width=.75in]{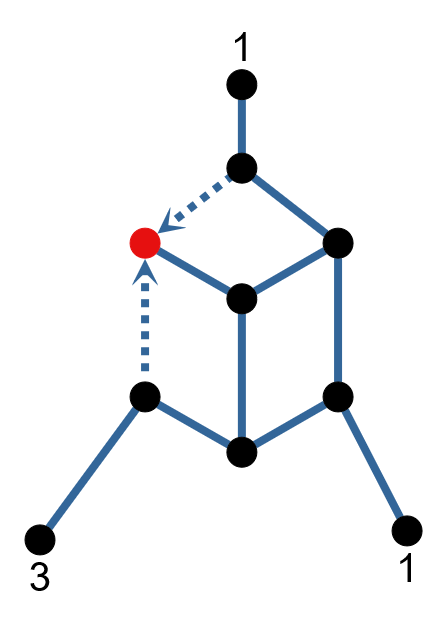}
\caption{
\label{fig: ExerciseSemidirected} 
The semi-directed network referenced in Exercise
\ref{exer: not phylogenetic}.
}
\end{center}
\end{figure}

\end{exercise}

\begin{exercise}
Find a formula for the reticulation number of 
a phylogenetic semi-directed network in terms 
of the number of leaves and edges of the network.
\end{exercise} 

\begin{exercise}
Consider the semi-directed topology
of the rooted 5-leaf network in
Figure \ref{fig: TwoPhylogeneticNetworks}.
How many different rooted phylogenetic networks
share this semi-directed topology?
\end{exercise}

\chproblem{
Prove or provide a counterexample to the following
statement.
It is impossible for two distinct 
phylogenetic semi-directed networks to have the same 
unrooted topology and the same set of reticulation
vertices (i.e., 
to differ only by which edges are the reticulation
edges).}

As a hint for this challenge problem,
consider the 3-leaf phylogenetic semi-directed 
network in Figure
\ref{fig: 3SemidirectedNetworks}.
Two of the reticulation
vertices are incident to leaf edges
in the network. As a first step, 
it may be helpful to consider whether or not
there is any way to reorient the edges into one of these
vertices so that it is still a reticulation 
vertex and so that the network remains a phylogenetic
semi-directed network.

\chproblem{Find an explicit formula for the number of  semi-directed networks with a single reticulation vertex and $n$ leaves.}

\resproject{
Find an explicit formula for the number of level-1 semi-directed networks with $n$ leaves and $m$ reticulation vertices. Can you generalize this formula to level-$k$ networks with $n$ leaves and $m$ reticulation vertices?
}

For Research Project 1, the level
of a semi-directed network is defined 
in terms of the unrooted, undirected topology just 
as for phylogenetic networks. Thus,
any phylogenetic network and its semi-directed 
topology will have the same level.  A starting point would be to look for patterns in small families of trees or networks. To begin thinking about proof techniques you might examine the proofs of the number of rooted trees with $n$ leaves, or perhaps the number of distinct unlabeled tree topologies with $n$ leaves.  Chapter three of Felsenstein's book \emph{Inferring Phylogeneies} provides some intuition about tree counting \protect\cite{Felsenstein2004}.

\chproblem{
Find necessary and sufficient
conditions for a semi-directed network
to be a phylogenetic semi-directed network.
}

\resproject{
Determine a method or algorithm 
for counting valid root locations
in a phylogenetic semi-directed network
(i.e., count the number of rooted
networks corresponding to a particular
semi-directed network).  
}

By definition, a phylogenetic semi-directed network must 
have at least one valid root location. A simple, though
extremely inefficient algorithm for finding all valid 
root locations would be to check all edges as root
locations and then all possible orientations for the other
edges. To improve on this naive algorithm, 
you might start by considering each reticulation
vertex one at a time. Does a single pair of reticulation
edges place restrictions on the possible valid root locations? 

It also might be helpful to have an 
efficient representation of a phylogenetic semi-directed
network. Since a phylogenetic semi-directed network is 
just a special type of graph, it can be represented by an
\emph{adjacency matrix}. There are some subtleties
involved in constructing this matrix for a semi-directed network, as there are both directed and 
undirected edges. Still, it could be useful to construct a dictionary between
properties of the network and properties of the
adjacency matrix of the network.

\resproject{Construct a fast heuristic algorithm which will determine if a semi-directed network is a phylogenetic semi-directed network. 
Alternatively, determine the computational complexity of determining if a given semi-directed network with $n$ leaves and reticulation number $m$ is a phylogenetic semi-directed network.}

These research projects may be closely related to 
Research Project 2 above. After all, determining 
if a semi-directed network is phylogenetic amounts
to determining if there exist \emph{any} valid
root locations. Thus, one might consider some 
of the suggestions above when approaching these
problems. Determining the computational complexity 
may prove very difficult indeed, and it may be a challenge
to prove something even when $m = 1.$ 

One general strategy for proving computational complexity results is to find a transformation from the problem of interest into another problem with a known computational complexity.  A good model for how this might work in the context of phylogenetics can be found in \protect\cite{anaya2016determining} a project which was the result of collaboration between undergraduates and faculty members.

\subsection{Restrictions of Networks}
In phylogenetics it is frequently necessary to pass back and forth between analyzing full datasets on a complete set of organisms $[n]$ and a more confined analysis on subset of $[n]$.  For instance you may have access to an existing data set on $[n]$ but are only interested in some subset of the organisms.  Alternatively you may have information on a collection of subsets of $[n]$ and want to piece them together to determine information about the complete set of organisms.  

 \begin{definition}
 \label{def: Restriction}
Let $N$ be an $n$-leaf phylogenetic network with root $\rho$,
and let $A \subseteq [n]$. The \emph{restriction of $N$
to $A$} is the phylogenetic network $N_{|A}$ constructed by
\begin{enumerate}[(i)]
\item Taking the union of all directed paths from 
$\rho$ to a leaf labeled by an element of $A$.
\item Deleting any vertices that lie above a vertex that is on every such path.
\item Suppressing all degree two vertices other than the root.
\item Removing all parallel edges.
\item Applying steps (iii) and (iv) until the network is a phylogenetic network.
\end{enumerate}

We say that $N$ \emph{displays} $N_{|A}$.
\end{definition}

\begin{figure}[h]
	  \caption{The restriction of a 6-leaf phylogenetic network to the set $\{1,4,5\}$.
	  The networks pictured are obtained by applying (i), (ii), and then (iii), (iv), and (iii) again to obtain the restricted phylogenetic network.}
	\label{fig: RestrictedNetwork} 
	\begin{center}
		\includegraphics[width=12cm]{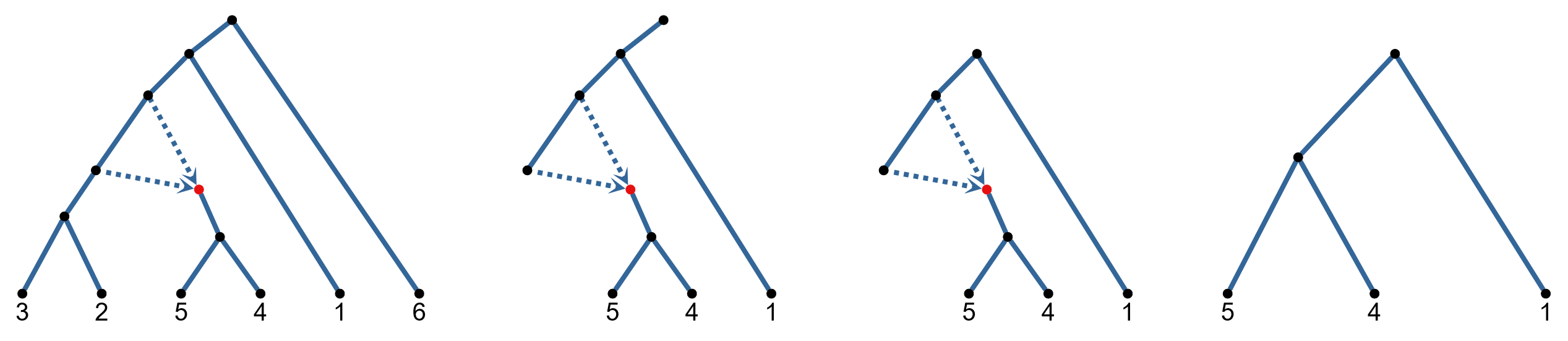}
		\end{center}
\end{figure}

While the definition of restriction
is defined in terms of a rooted phylogenetic network,
we can also apply this definition
to a semi-directed network. Given
a phylogenetic semi-directed network, its restriction to a subset $A \subset [n]$ is found by
rooting the network at a valid root
location, restricting the rooted
phylogenetic network to $A$, and 
then taking the semi-directed topology
of the restricted phylogenetic network. The following Challenge Problem shows that this
operation is well-defined.

\chproblem{
Suppose that a valid rooting
is chosen for an $n$-leaf semi-directed network and that the network is then restricted to a subset of the leaves
of size $k \leq n$. 
Show that the $k$-leaf semi-directed
network obtained by unrooting the
restricted network is independent of
the original rooting chosen.}

In practice it can be computationally difficult to directly estimate a phylogenetic network from sequence data corresponding to the set $[n]$. One potential workaround is to infer phylogenetic networks on a collection of subsets of $[n]$, and then select a larger network $N$ which best reflects the networks estimated on the various subsets.

\begin{definition}
A set of phylogenetic networks $\mathcal{A}=\{N_1,N_2,\cdots, N_k\}$ whose leaves are all contained in a set $[n]$ is called \emph{compatible} if there exists a phylogenetic network $N$ for which the restriction of $N$ to the leaf set of $N_i$ is isomorphic to $N_i$ for all $1\le i \le k$.  
\end{definition}

It is common when working with unrooted
trees to restrict the trees to four
element subsets of the leaves.
The resulting 4-leaf trees are called 
\emph{quartets}, and an $n$-leaf phylogenetic tree is uniquely determined by its $n \choose 4$ quartets. Similarly, when working with a network,
we can construct a \emph{quarnet} by restricting
the network to a four element subset of its leaves. In this paper,
since we are working with semi-directed phylogenetic networks, we will use the term
quarnet to mean a 4-leaf semi-directed phylogenetic network. However, note that in other sources a quarnet may refer to an unrooted 4-leaf network.

\begin{exercise}
Determine if the following collections of quarnets are compatible. 

\begin{itemize}
    \item[(a)]  
        \begin{center}
            \includegraphics[width =10cm]{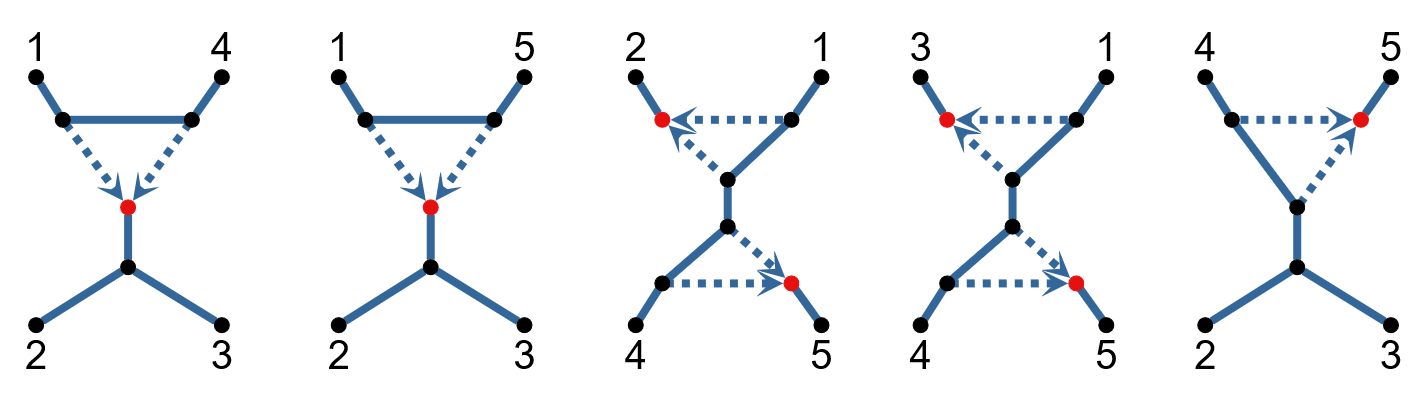}
        \end{center}
    \item[(b)]  
        \begin{center}
            \includegraphics[height =2.75cm]{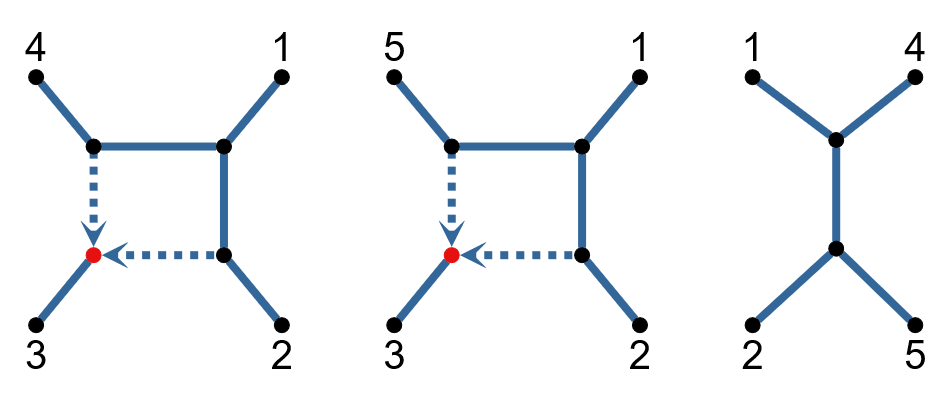}
        \end{center}
    \end{itemize}
\end{exercise}

It is possible that some collections of subnetworks can be displayed by multiple phylogenetic networks.  In practice we might want to know when a collection of subnetworks can be used to represent a unique network.

\begin{definition}
Let $\mathcal{A}=\{N_1,N_2,\cdots, N_k\}$ be a collection of phylogenetic networks for which the union of all of the corresponding leaf sets is $[n]$. The collection $\mathcal{A}$  is said to \emph{distinguish} a phylogenetic network $N$, if $N$ is the only phylogenetic network with leaf set $[n]$ such that the restriction of $N$ to the leaf set of $N_i$ is isomorphic to $N_i$ for all $1\le i \le k$.
\end{definition}

\begin{exercise}
Find a collection of three quarnets which are displayed by the 6-leaf phylogenetic network in Figure~\ref{fig: RestrictedNetwork} which do not distinguish that network.
\end{exercise}

\chproblem{
Show that the set of all
quarnets of a level-one semi-directed network distinguishes that network.
}

A good strategy for proving this
might be to consider two distinct
level-one semi-directed networks, 
and then show that there
must be a quarnet on which they 
differ. 

\chproblem{Find all minimal sets of quarnets which distinguish the semi-directed topology of the 6-leaf phylogenetic network in Figure~\ref{fig: RestrictedNetwork}.}

\chproblem{Give criteria for determining whether or not a collection of quarnets are compatible. Hint: There are known criteria for determining if a set of quartet trees are compatible \cite{grunewald2008quartet}.}

\resproject{
Describe an algorithm that
determines if a set of semi-directed networks $\mathcal{A}=\{N_1,N_2, \cdots, N_k\}$ is compatible.
Bonus points if the algorithm is efficient, constructive, or determines if the collection distinguishes a unique network.  This question is already interesting in the case that each of the $N_i$ is a quarnet.}

The previous research problem is based on the notion that one could computationally estimate quarnets from DNA-sequence data, and then the compatible quarnets could be combined to determine a single network which describes the evolution across a broader collection of organisms. 
This idea has proven successful when building
phylogenetic trees, thus, a number of authors have
studied whether or not networks can be constructed by building up large networks from smaller structures (e.g.,
\cite{Huber2018Quarnet,VanIersel2014,
Jansson2006, Keijsper2013}).
Insights and techniques from these
papers will likely prove valuable for attacking some of these research and challenge questions. 
However, it is unlikely that any results will translate
directly, since each of the sources cited place different
restrictions on the types of input networks and the types
of networks constructed.

As a warmup to this activity one might examine similar results on trees as can be found throughout Chapters 3 and 6 in the textbook \emph{Basic Phylogenetic Combinatorics} \protect\cite{dress2012basic}, the introductory chapters of which also provide a nice mathematical framework for working with trees and networks.  However there is a level of abstraction in this book which mandates that readers may need to keep a running list of concrete examples nearby to connect the text with their intuitive understanding of trees and networks.

\begin{exercise}
Construct a phylogenetic network which displays the following quarnets:  Either prove this collection distinguishes the network, or find the set of all networks which display this collection. 
\end{exercise}

\begin{center}
\includegraphics[width =10cm]{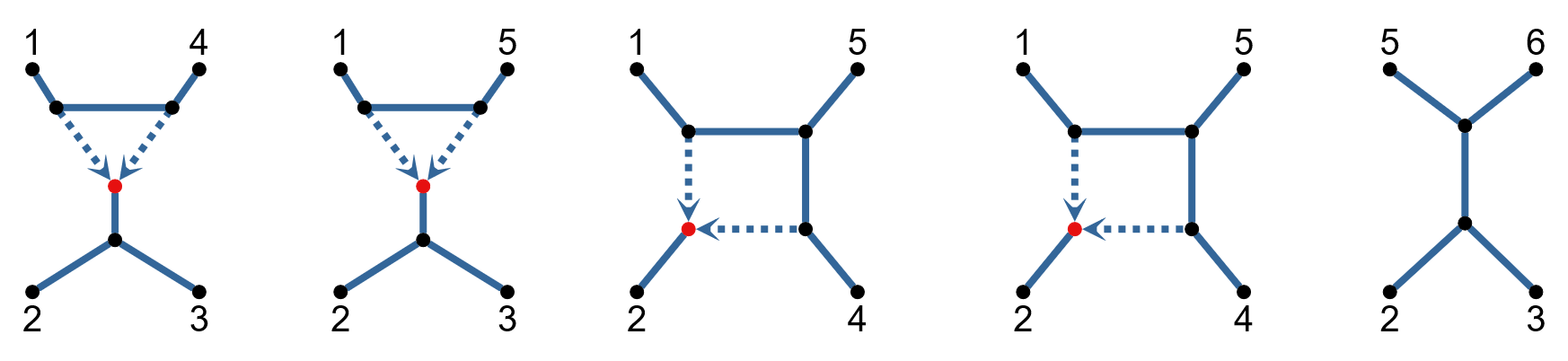}
\end{center}

In practice, the estimation of quarnets from data is likely to be imperfect. Thus, even if we produce data from a model on an $[n]$-leaf semi-directed network, the collection of estimated quarnets is likely to be incompatible.
The same issue applies no matter the size
of the the inferred subnetworks.
In such cases, one would like to construct a phylogenetic network which displays the maximum number of quarnets or other semi-directed networks in a collection $\mathcal{A}$.

\begin{exercise}
\label{exer: LargeQuarnetSet}
Find a phylogenetic network which displays the maximum number of the following collection of quarnets. 
\end{exercise}

\begin{center}
\includegraphics[width =8cm]{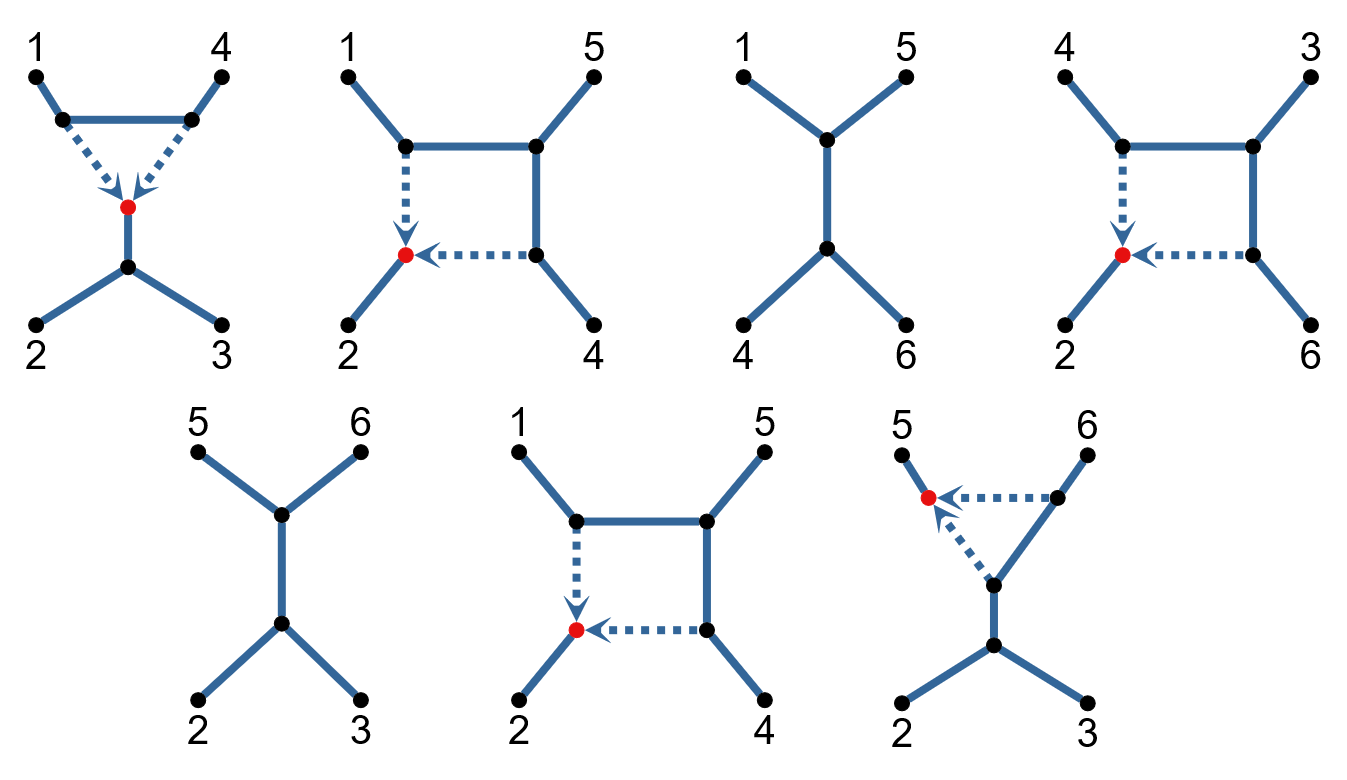}
\end{center}

\resproject{Given a collection $\mathcal{A}=\{N_1,N_2,\cdots, N_k\}$ of semi-directed networks, give an efficient algorithm for computing a semi-directed network $N$ which maximizes the number of sub-networks in $\mathcal{A}$ displayed by $N$.}

As a warmup example, in the case of trees, there are two very popular approaches to this problem. The first is described in a series of papers describing the  ASTRAL family of software \protect\cite{mirarab2014astral}, where the solution tree $T$ is assumed to have certain features which appear in the trees in $\mathcal{A}$.  This is a very efficient algorithm which provably solves the problem under this assumption.  An alternative is the quartet MaxCut family of algorithms \protect\cite{snir2012quartet} which provide a fast heuristic for solving this problem.  While it does not offer the same theoretical guarantees of the ASTRAL methods, it also removes some of the restrictive assumptions of
the ASTRAL method.  Both of these algorithms would be good starting points for exploration.

In moving towards networks, one might examine the recent software SNAQ \protect\cite{Solis2016} which builds phylogenetic networks based on input from a collection of quarnets.  This research problem is very broad, and allows for many types of restrictions that would still be interesting in practice.  One should feel free to restrict both the types of networks in the collection $\mathcal{A}$ and the type of semi-directed network $N$ which is allowed.  Consider restrictions both on the number of leaves, 
level, and number of reticulation vertices.

\section{Algebra of Phylogenetic Networks}
\label{sec: Algebra of Phylogenetic Networks}

In the previous section, we introduced networks
as a possible explanation for the evolutionary
history of a set of species and explored 
some combinatorial questions related to their structure. In this section we move from combinatorial to algebraic questions. In particular, we study \emph{phylogenetic ideals}, collections of polynomials associated to 
models of DNA sequence evolution.  Phylogenetic ideals associated to tree models have been well-studied (e.g., \cite{Allman2006,  Eriksson2009, Sturmfels2006}) and have been used not only for model selection but also to prove theoretical results about the  models.  For example, they have been used to show that 
the tree parameters of certain models are \emph{identifiable} (e.g., \cite{ Allman, Allman2006, Chang1996, Long2015a}). 
A model parameter is identifiable if each output from the model uniquely determines the value of that
parameter.
This is an important consideration for using 
phylogenetic models for inference, since it would
be undesirable to have multiple different trees
explain our data equally well.

Phylogenetic ideals are determined by two things: a model of DNA sequence evolution and a tree or network. 
In this section, we fix the model of DNA sequence evolution, and then focus on how the polynomials change based on different network attributes.  
The model of evolution that is quietly sitting in the background is the Cavendar-Farris-Neyman (CFN) model. While there are four DNA bases
(adenine ($A$), guanine ($G$), cytosine ($C$), and thymine ($T$)), the CFN model only distinguishes between purines ($A, G$) and pyrimidines ($C,T$).
Thus, it is a 2-state model of evolution
where the two states are represented by $0$ and $1$.  For the CFN model on a fixed $n$-leaf tree $T$, 
the mutations between purines and pyrimidines are modeled as a Markov process proceeding along the tree.
The numerical parameters of the model determine
the probabilities that mutations occur along
each edge. 
Once the numerical parameters are specified, the model gives a probability distribution on the set $\{0,1\}^n$. Put another way, the tree
determines a map,
or \emph{parameterization}, that sends each choice
of numerical parameters to a probability distribution. 
Because each coordinate of this map is a polynomial, 
we can consider it as a ring homomorphism.
The kernel of this homomorphism is the phylogenetic ideal associated to $T$.

In this section, we describe how to associate
an ideal to a phylogenetic semi-directed network.
Just as for trees, we begin by describing
how to construct a polynomial map from the network,
and the ideal of the network is the kernel of this map.
We then present a number
of Research Projects related to uncovering generating sets and properties of these ideals as well as comparing the ideals for different networks.
Phylogenetic network ideals were originally studied in \cite{Gross2017distinguishing}, and it is likely
they will receive increasing attention as researchers look to apply methods that have proven successful for trees to phylogenetic networks.

While not essential for the projects
presented below, for those interested in learning
more about the CFN model and the connections to 
phylogenetic ideals we recommend \cite{Gascuel2007}.
One reason that we do not dwell on the details of
the maps referenced above is that we actually work in a set of transformed coordinates called the \emph{Fourier coordinates}, introduced in \cite{Evans1993}.
This is common when studying phylogenetic ideals, as it makes
many of the computations feasible.
Though the derivation and details of the transform are outside the scope of this chapter, they can be found in \cite{Eriksson2009, Evans1993, Sturmfels2005}.
Viewing phylogenetic statistical models from an algebraic perspective fits broadly into the field
of \emph{algebraic statistics}. An overview
of some of the basic concepts and significant
results in this area can be found in \cite[Chapter 15]{Sullivant2018algebraic}.
Similarly, many of the concepts below come from
computational algebraic geometry, and some good
first references for students are \cite{Cox2007, Hassett2007}. If the reader has not yet had a course in abstract algebra, \cite[Chapter 4]{Allman2004Mathematical} provides an excellent introduction to the algebraic viewpoint on phylogenetics which is accessible to readers who are familiar with matrices.

\subsection{Ideals Associated to Sunlet Networks} 
\label{sec: sunlet}

The algebra of phylogenetic semi-directed networks is rich enough that even the simplest  networks give rise to interesting research questions. Therefore, in this chapter, we will work with semi-directed networks with only a single reticulation vertex.
As an undirected graph, a semi-directed network
with a single reticulation has a unique cycle of length $k$, and so we call these
semi-directed networks \emph{$k$-cycle networks}. To begin this section, we will first consider a
specific type of $k$-cycle network called a \emph{sunlet} network. A \emph{$k$-sunlet} network is a $k$-leaf, $k$-cycle network. 
Starting with sunlet networks will allow us to 
introduce network ideals in a simplified setting, before we show how to associate an ideal
to a general $k$-cycle network in Section \ref{sec: beyondsunlets}.

\begin{figure}[h]
	  \caption{A 5-cycle network and a 6-sunlet.}
	\label{fig: 2kcyclenetworks} 
	\begin{center}
		\includegraphics[width=8cm]{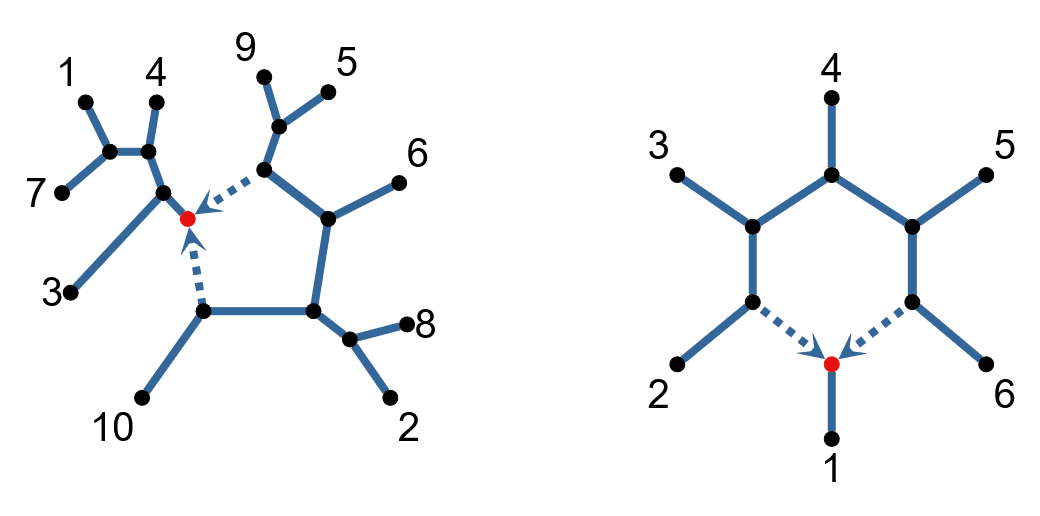}
		\end{center}
\end{figure}

Let $N_k$ be the $k$-sunlet network with the leaves labeled clockwise, from 1 to $k$, starting from the leaf extending from the single reticulation vertex. For example, $N_6$ is
the network pictured at right in Figure
\ref{fig: 2kcyclenetworks}.
In order to describe the ideal associated
to $N_k$ we will need to introduce two polynomial rings. 
For what follows, we will use $\mathbb Z_2$ to denote the quotient group $\mathbb {Z} /2\mathbb {Z}$.  This group has two elements, $0$ and $1$, with addition modulo 2.
The first polynomial ring we will consider is

$$R_k:=\mathbb{Q}[q_{i_1, \ldots, i_k} \ : \ i_1, \ldots, i_k \in \mathbb{Z}_2, \ i_1 + \ldots + i_k = 0]$$

\begin{exercise} Enumerate the indeterminates, i.e. variables, for $R_3$. In general, how many indeterminates does $R_k$ have?\end{exercise}

\begin{figure}[h]
	  \caption{The 4-sunlet network $N_4$.}
	\label{fig: 4sunlet} 
	\begin{center}
		\includegraphics[width=2.5cm]{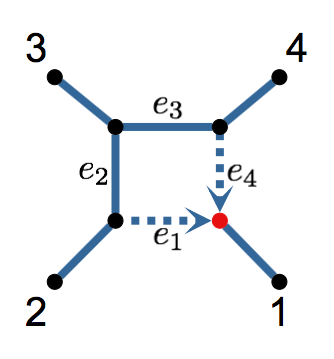}
		\end{center}
\end{figure}

The next ring we will consider is a ring
with two indeterminates associated to each edge in $N_k$. The $k$-sunlet network $N_k$ has $2k$ edges, $k$ of which are leaf edges and $k$ of which
are internal (non-leaf) edges.
We label the leaf edges of the network from $1$ to $k$ to match the corresponding
leaf labels. Similarly, we label the internal edges from $1$ to $k$, starting
with the reticulaton edge clockwise from the leaf edge labeled by $1$ and 
continuing around the sunlet
(as in Figure \ref{fig: 4sunlet}).
To each edge of the sunlet, we associate two indeterminates, one for each element of $\mathbb Z_2$.
We denote the indeterminates for the leaf edge labeled by $i$ as $a^{(i)}_0$ and $a^{(i)}_1$  and the indeterminates for the internal edge labeled by $i$ as
$b^{(i)}_0$ and $b^{(i)}_1$.
The second polynomial ring of interest is 
$$S_k:=\mathbb{Q}[a^{(i)}_{j}, \ b^{(i)}_{j} \ : 1 \leq i \leq k, \ j \in \mathbb{Z}_2]$$

An ideal $I$ of a ring $R$ is a subset of $R$ closed under addition and multiplication by ring elements, that is, for all $f, g \in I$, we have $f+g \in I$, and for all $r \in R$ and $f \in I$, we have $rf \in I$. The ideal $I_k$ associated to the phylogenetic network $N_k$ is the kernel of the following ring homomorphism:
\[
\phi_k: R_k \to S_k \]
\[q_{i_1, \ldots, i_k} \mapsto a^{(1)}_{i_1} \cdots a^{(k)}_{i_k}( \prod_{j=1}^{k-1} b^{(j)}_{i_1 + \ldots + i_j} + \prod_{j=2}^{k} b^{(j)}_{i_2 + \ldots + i_j}).
\]
In other words, 
$$I_k := \ker(\phi_k)= 
\{f \in R_k \ : \ \phi_k(f) = 0\}.$$

\begin{exercise}
Show that the kernel of any ring homomorphism
is an ideal.
\end{exercise}

\begin{exercise}
For $k=3$ write down the rings $R_k$ and $S_k$. Let $f=3q_{1,1,0}q_{1,0,1}^2+q_{0,0,0}$.  Compute $\phi_{3}(f)$.
\end{exercise}

\begin{exercise}
Find a non-zero polynomial in $I_3$ or prove that no such polynomial exists.
\end{exercise}

The ideal $I_{k} \subseteq R_k$ is finitely generated, meaning that there exist  $g_1, \ldots, g_m \in R_k$ such that for any $f \in I_{k}$, there exist $r_1, \ldots, r_m \in R_k$ such that 
$f = r_1g_1 + r_2g_2 + \ldots + r_mg_m$. 
Any set $\{g_1, \ldots, g_m\}$ that satisfies the preceding definition is called a \emph{generating set} of $I_{k}$. When studying ideals associated to phylogenetic networks, we are interested in the polynomials in the ideal. In some cases, just knowing a few polynomials in the ideal is helpful, but we can obtain a more complete understanding of the ideal if we can determine a generating set.

There are algorithms based on the theory of Gr\"{o}bner bases for determining the
generating set for an ideal from its parameterization.
A Gr\"{o}bner basis is a
special type of generating set 
for an ideal and we encourage curious readers to learn more about them before starting on some of the research problems in this section (see e.g. \cite{Cox2007, sturmfels1996grobner}).
However, while these algorithms give us a means of determining a generating set for an ideal, in most cases of interest, it is infeasible to perform all the computations necessary by hand. Therefore, we will want to use a computer algebra system to do most of the tedious work for us. 
In this chapter, we will use the computer algebra system \emph{Macaulay2} \cite{M2}. 
As a first example, we show below how
to use this program to find a generating set for $I_4$, the ideal associated to the 4-sunlet network.

\begin{example}
\label{ex: sunlet ideal}
Let us consider $N_4$, the $4$-leaf sunlet network pictured in Figure \ref{fig: 4sunlet}.  In this case, the two rings of interest are
$$R_4 = \mathbb Q[q_{0000}, q_{0011}, q_{0101}, q_{0110},q_{1001},q_{1010},q_{1100}, q_{1111}], \text{ and }$$
$$S_4 = \mathbb{Q}[a_0^{(1)},a_1^{(1)},a_0^{(2)},a_1^{(2)},a_0^{(3)},a_1^{(3)},a_0^{(4)},a_1^{(4)},b_0^{(1)},b_1^{(1)},b_0^{(2)},b_1^{(2)},b_0^{(3)},b_1^{(3)},b_0^{(4)},b_1^{(4)}].$$
The ring homomorphism $\phi_4$ is described as follows:
\begin{align*}
\phi_{4}(q_{0000}) &=\ a_0^{(1)}a_0^{(2)}a_0^{(3)}a_0^{(4)}\big(b_0^{(1)}b_0^{(2)}b_0^{(3)} + b_0^{(2)}b_0^{(3)}b_0^{(4)}\big), \\ \phi_{4}(q_{0011}) &=\ a_0^{(1)}a_0^{(2)}a_1^{(3)}a_1^{(4)}\big(b_0^{(1)}b_0^{(2)}b_1^{(3)} + b_0^{(2)}b_1^{(3)}b_0^{(4)}\big), \\
\phi_{4}(q_{0101}) &=\ a_0^{(1)}a_1^{(2)}a_0^{(3)}a_1^{(4)}\big(b_0^{(1)}b_1^{(2)}b_1^{(3)} + b_1^{(2)}b_1^{(3)}b_0^{(4)}\big), \\
\phi_{4}(q_{0110}) &=\ a_0^{(1)}a_1^{(2)}a_1^{(3)}a_0^{(4)}\big(b_0^{(1)}b_1^{(2)}b_0^{(3)} + b_1^{(2)}b_0^{(3)}b_0^{(4)}\big), \\
\phi_{4}(q_{1001}) &=\ a_1^{(1)}a_0^{(2)}a_0^{(3)}a_1^{(4)}\big(b_1^{(1)}b_1^{(2)}b_1^{(3)} + b_0^{(2)}b_0^{(3)}b_1^{(4)}\big),
\\
\phi_{4}(q_{1010}) &=\ a_1^{(1)}a_0^{(2)}a_1^{(3)}a_0^{(4)}\big(b_1^{(1)}b_1^{(2)}b_0^{(3)} + b_0^{(2)}b_1^{(3)}b_1^{(4)}\big),
\\
\phi_{4}(q_{1100}) &=\ a_1^{(1)}a_1^{(2)}a_0^{(3)}a_0^{(4)}\big(b_1^{(1)}b_0^{(2)}b_0^{(3)} + b_1^{(2)}b_1^{(3)}b_1^{(4)}\big),\\
\phi_{4}(q_{1111}) &=\ a_1^{(1)}a_1^{(2)}a_1^{(3)}a_1^{(4)}\big(b_1^{(1)}b_0^{(2)}b_1^{(3)} + b_1^{(2)}b_0^{(3)}b_1^{(4)}\big).
\end{align*}
Using \emph{Macaulay2} we can compute a generating set for $I_{4}$. 
In the code below, we use {\tt{I}} for this ideal, {\tt{R}} and {\tt{S}} for the rings $R_4$ and $S_4$, and {\tt{phi}} for the map $\phi_{4}$.

{\small
\begin{verbatim}
i1: R   = QQ[q_{0,0,0,0}, q_{0,0,1,1}, q_{0,1,0,1}, q_{0,1,1,0}, 
             q_{1,0,0,1}, q_{1,0,1,0}, q_{1,1,0,0}, q_{1,1,1,1}];

i2: S   = QQ[a1_0, a1_1, a2_0, a2_1, a3_0, a3_1, a4_0, a4_1, 
             b1_0, b1_1, b2_0, b2_1, b3_0, b3_1, b4_0, b4_1];

i3: phi = map(S, R, 
      {a1_0*a2_0*a3_0*a4_0*(b1_0*b2_0*b3_0+b2_0*b3_0*b4_0),
       a1_0*a2_0*a3_1*a4_1*(b1_0*b2_0*b3_1+b2_0*b3_1*b4_0),
       a1_0*a2_1*a3_0*a4_1*(b1_0*b2_1*b3_1+b2_1*b3_1*b4_0),
       a1_0*a2_1*a3_1*a4_0*(b1_0*b2_1*b3_0+b2_1*b3_0*b4_0),
       a1_1*a2_0*a3_0*a4_1*(b1_1*b2_1*b3_1+b2_0*b3_0*b4_1),
       a1_1*a2_0*a3_1*a4_0*(b1_1*b2_1*b3_0+b2_0*b3_1*b4_1),
       a1_1*a2_1*a3_0*a4_0*(b1_1*b2_0*b3_0+b2_1*b3_1*b4_1),
       a1_1*a2_1*a3_1*a4_1*(b1_1*b2_0*b3_1+b2_1*b3_0*b4_1)})

i4: I   = ker phi

o4: ideal(q_{0, 1, 1, 0}*q_{1, 0, 0, 1}-q_{0, 1, 0, 1}*q_{1, 0, 1,0}+
          q_{0, 0, 1, 1}*q_{1, 1, 0, 0}-q_{0, 0, 0, 0}*q_{1, 1, 1, 1})
\end{verbatim} }

The output of the last command tells us that $I_4$ is generated by a single polynomial, namely 
$$q_{0110}q_{1001} - q_{0101}q_{1010} + q_{0011}q_{1100} - q_{0000}q_{1111}.$$
\end{example}

\begin{exercise}\label{ex:verifypolynomial} 
Verify that the polynomial from 
Example \ref{ex: sunlet ideal} is in $I_{4}$.
\end{exercise}

\begin{exercise} \label{exer: I5 computation}
Compute the ideal $I_5$ for the the 5-sunlet $N_5$ using \emph{Macaulay2} or another computer algebra system.  How many generators are returned?  What are the degrees of the returned generators? 
\end{exercise}

\begin{exercise} \label{exer: lifting}
Verify (computationally or by hand) that the polynomial
$$q_{01100}q_{10010} - q_{01010}q_{10100} + q_{00110}q_{11000} - q_{00000}q_{11110}$$
is in the ideal $I_5$.
\end{exercise}

On a standard laptop, the computation in Exercise \ref{exer: I5 computation} will finish, 
but not immediately. 
You may notice the difference in the time
it takes to run the computation
for $I_4$ in Example \ref{ex: sunlet ideal} 
and for $I_5$ in Exercise \ref{exer: I5 computation}.  
As we increase $k$, 
computing $I_k$ becomes even more complex,
to the point that a computer may take several hours or days or may run out of memory before returning
a generating set.
The computer, of course, will execute an algorithm to determine a generating set for $I_k$.
In many cases however, executing all the steps
of the algorithm is not actually necessary to obtain the information about the ideal that we are interested in. Therefore, we can use some tricks
and techniques to reduce the size of the computations and extract information about the ideals.

For example, we can use some of the built-in options in \emph{Macaulay2} such as {\tt SubringLimit}, a command that stops the computation after a specified number of polynomials have been found.
If using this strategy, we will obtain a set
of polynomials in the ideal $I_k$, but we will not
have a certification that these polynomials generate $I_k$. However, if we let $J$ be the ideal they generate then we know that $J \subseteq I_k$. 
We can show that $J = I_k$ if we can show that $J$ is  \emph{prime} and that the \emph{dimension} of $J$ is equal to that of $I_k.$
An ideal $I \subseteq R$ is prime if for all $f,g \in R$, if $fg \in I$, then $f \in I$ or $g \in I$. 
Checking whether an ideal is prime and finding its dimension can be done in 
\emph{Macaulay2} using
the {\tt isPrime} and {\tt dim} commands.
Of course, we do not have a set of generators for $I_k$, since that is what we are trying to find, so we can not use {\tt dim} to find its dimension. However, we can still determine a lower bound
on the dimension of $I_k$ from the 
map $\phi_k$ using the rank of the
Jacobian matrix as shown in Example~\ref{rankandprime}.
Since $J \subseteq I_k$, we have $\dim(I_k) \leq \dim(J)$, and so if the rank of the Jacobian is equal to $\dim(J)$, then
$\dim(I_k) = \dim(J)$.

This {\tt SubringLimit} method of determining a generating set for an ideal was used to prove Proposition 4.6 in \cite{Gross2017distinguishing}. 
That paper also includes
supplementary \emph{Macaulay2} code which may prove useful.

\begin{example}
\label{rankandprime}
Let $I=\langle q_{0110}q_{1001} - q_{0101}q_{1010} + q_{0011}q_{1100} - q_{0000}q_{1111} \rangle$ be the ideal returned from Example \ref{ex: sunlet ideal}. The following \emph{Macaualy2} code is used to determine whether the dimension of the ideal $I$ is the same as the dimension of the ideal $I_4$ as well as whether or not $I$ is prime.  This serves as verification that $I$ is indeed equal to $I_4$.

{\small
\begin{verbatim}

i5: phimatrix = matrix{{
      a1_0*a2_0*a3_0*a4_0*(b1_0*b2_0*b3_0+b2_0*b3_0*b4_0),
       a1_0*a2_0*a3_1*a4_1*(b1_0*b2_0*b3_1+b2_0*b3_1*b4_0),
       a1_0*a2_1*a3_0*a4_1*(b1_0*b2_1*b3_1+b2_1*b3_1*b4_0),
       a1_0*a2_1*a3_1*a4_0*(b1_0*b2_1*b3_0+b2_1*b3_0*b4_0),
       a1_1*a2_0*a3_0*a4_1*(b1_1*b2_1*b3_1+b2_0*b3_0*b4_1),
       a1_1*a2_0*a3_1*a4_0*(b1_1*b2_1*b3_0+b2_0*b3_1*b4_1),
       a1_1*a2_1*a3_0*a4_0*(b1_1*b2_0*b3_0+b2_1*b3_1*b4_1),
       a1_1*a2_1*a3_1*a4_1*(b1_1*b2_0*b3_1+b2_1*b3_0*b4_1)}}

i6: rank jacobian phimatrix == dim(I)

i7: isPrime I

\end{verbatim} }
\end{example}

\chproblem{Compute $I_6$ in \emph{Macaulay2} by imposing a limit on the number of polynomials returned using {\tt SubringLimit}. Verify that the ideal that is returned is indeed $I_6$. }

One will only get so far using the strategy described above,
as for larger $k$, there may be many polynomials required to generate $I_k$ and they may take a very long time to find. In these cases, just being able to compute the ideal $I_k$ becomes an interesting project on its own.

\resproject{
Find a generating set for the ideal $I_k$
of the $k$-sunlet network $N_k$ when $k=7,8,9$.}

Moving from the computational to the theoretical, 
it is sometimes possible to give a 
description of a generating set for a whole
class of ideals.

\resproject{Give a description of a set of phylogenetic invariants in the sunlet ideal $I_k$.  Does this set of invariants generate the ideal? Does this set of invariants form a Gr\"{o}bner basis for the ideal with respect to some term order?}

We can envision two different approaches to Research Project 7.  The first is to compute the sunlet ideals for a range of examples. As you are able to compute $I_k$ for higher $k$, patterns should emerge. 
We see this even for $k=4$ and $k=5$.
For example, Exercise \ref{exer: lifting} 
might give a hint of how we can find some invariants for larger $k$ by doing computations for small $k$.
Once you discover a pattern, you could then try to prove that this
pattern holds in general.

The second approach would be to try to construct invariants for sunlet networks using the known invariants in the ideals of the trees that they display. 
As an example, consider the map $\phi_4$ from Example \ref{ex: sunlet ideal}. 
Consider the following two maps. 
The first, $\phi': R_4 \to S_4$, sends $q_{i_1i_2i_3 i_4}$ to the term of $\phi_4(q_{i_1i_2i_3 i_4})$ that includes $b^{(1)}_j$ and the other, 
$\phi'': R_4 \to S_4$, sends
$q_{i_1i_2i_3 i_4}$ to the second term of $\phi_4(q_{i_1i_2i_3 i_4})$
that includes $b^{(4)}_j$.
So, for example,
$$\phi'(q_{0000}) = a_0^{(1)}a_0^{(2)}a_0^{(3)}a_0^{(4)}b_0^{(1)}b_0^{(2)}b_0^{(3)} \text{ and } \phi^{''}(q_{0000}) =\ a_0^{(1)}a_0^{(2)}a_0^{(3)}a_0^{(4)}b_0^{(2)}b_0^{(3)}b_0^{(4)}. $$ 
The ideal $I' = \ker(\phi')$ is the ideal of the tree created by removing the reticulation edge $e_4$ from the 4-sunlet in Figure \ref{fig: 4sunlet}. 
The ideal $I'' = \ker(\phi'')$ is the ideal of the tree created by removing the reticulation edge $e_1$.

The problem of finding invariants for trees
has been solved, and for small trees, explicit lists of invariants are available online at \url{https://www.shsu.edu/~ldg005/small-trees/}, the work of which is described in Chapter 15 of \cite{Pachter2005}. 
The tree ideals are parametrized by monomials which makes it easier to find invariants. 
In particular, invariants for ideals parameterized by monomials can be found by examining the additive relationships between the exponents of the monomials.
This means that finding invariants for
these ideals can be done using only tools from linear algebra.

\begin{example}
\label{ex: toric}
Let $f: \mathbb{R}^2 \rightarrow \mathbb{R}^3$ be the map
defined by $(t_1,t_2) \mapsto (t_1^2,t_1t_2,t_2^2)$. 
We can represent
this map by a $2 \times 3$ matrix $A$,
where the $ij$-th entry is the exponent of $t_i$
in the $j$-th coordinate of the image of $(t_1, t_2)$,
$$A = 
\begin{pmatrix}
2 & 1 & 0 \\
0 & 1 & 2 \\
\end{pmatrix}.
$$
Elements of the integer kernel 
of $A$ encode binomial invariants in 
$\ker(f)$. 
For example, the integer vector
$(1,-2,1)^T$
is a vector of integers
in $\ker(A)$. 
We can interpret the positive entries
as the monomial $y_1y_3$
and the negative integers as the monomial $y_2^2$, and conclude that
$y^2 - y_1y_3$ is in $\ker(f)$.
\end{example}

Notice in the preceding example that while the parameterization was in terms of monomials, the invariant we 
constructed is a binomial.
While there are many different formal definitions, the class of ideals which are parameterized by monomials are called \emph{toric ideals} and it is 
known that toric ideals can be generated by binomials.
This fact is proven in 
\cite[Chapter 4]{sturmfels1996grobner},
which might also serve as a good reference for learning more about the invariants of toric ideals. 
The following exercise shows why toric ideals may prove useful when trying to find invariants for sunlet ideals.

\begin{exercise}

Consider the ideals $I_4$, $I' = \ker(\phi')$, and
$I'' = \ker(\phi'')$ described above.

\begin{itemize} 
    \item[(a)] \ \ Show that if $f \in I_4$, then $f \in I' \cap I''.$ 
    \item[(b)] \ \ Compute $I' = \ker(\phi')$ and
    $I'' = \ker(\phi'')$ using \emph{Macaulay2}. You can verify that your computations are correct using the online catalog of invariants referenced above. Specifically, by looking under ``Invariants in Fourier coordinates" for the ``Neyman 2-state model." 
    \item[(c)] \ \ Verify that the 
    generator for $I_4$ found in Example \ref{ex: sunlet ideal} is contained in $I'$ and in $I''$. (Hint: 
    To determine if a polynomial $f$ is contained in an ideal
    $I$, you can verify in \emph{Macaulay2} that {\tt f \% I == 0} 
    returns {\tt TRUE}). 
\end{itemize}

\end{exercise}

The previous exercise shows that $I_4 \subset I' \cap I''$. 
Put another way,
invariants in $I'$ and $I''$ are candidates
to be invariants in $I_4.$ 
Similar statements hold for all of the
ideals $I_k$ in this section, 
and for the ideals $I_N$ that we describe in the next section.  Thus, exploring toric ideals 
might prove useful for finding network invariants.

\subsection{Beyond sunlet networks} \label{sec: beyondsunlets}

The sunlet networks $N_k$ have a very particular structure, and the ring map we described in Section \ref{sec: sunlet} is specific to sunlets.
In this section, we set up the ring map $\phi$ more generally, which will allow us to explore the algebra of general $k$-cycle networks.  

Let $N$ be an $n$-leaf, $k$-cycle network.  The first ring we will consider is of the same form as that from the previous section,

$$R_n:=\mathbb{Q}[q_{i_1, \ldots, i_n} \ : \ i_1, \ldots, i_n \in \mathbb{Z}_2, \ i_1 + \ldots + i_n = 0].$$

The next ring we will consider is a ring
with two indeterminates associated to each edge of
 $N$. 
As with the sunlet, an $n$-leaf, $k$-cycle network has $2n$ edges, but unlike with sunlets, 
we no longer make a distinction between the leaf edges and the interior edges when labeling and so label all the edges by $\{1, \ldots, 2n\}$.
As before, we associate two parameters to each edge, indexed by the edge label and the elements of $\mathbb{Z}_2$.

$$S_{n}:=\mathbb{Q}[
a^{(i)}_{0}, a^{(i)}_{1} : 1 \leq i \leq 2n].$$

Our next step will be to define the map 
$\phi_N: R_n \to S_n$.
Before we do this, observe that if we remove one of the reticulation edges of ${N}$, the result is an unrooted $n$-leaf tree with
labeled leaves.
These two trees, $T_1$ and $T_2$, are not binary since removing a reticulation edge in the network will leave vertices of degree two.

\begin{figure}[h]
	  \caption{The 4-sunlet network $N_4$ and the two trees obtained by removing each reticulation edge.}
	\label{fig: 4sunletlabeled} 
	\begin{center}
		\includegraphics[width=8cm]{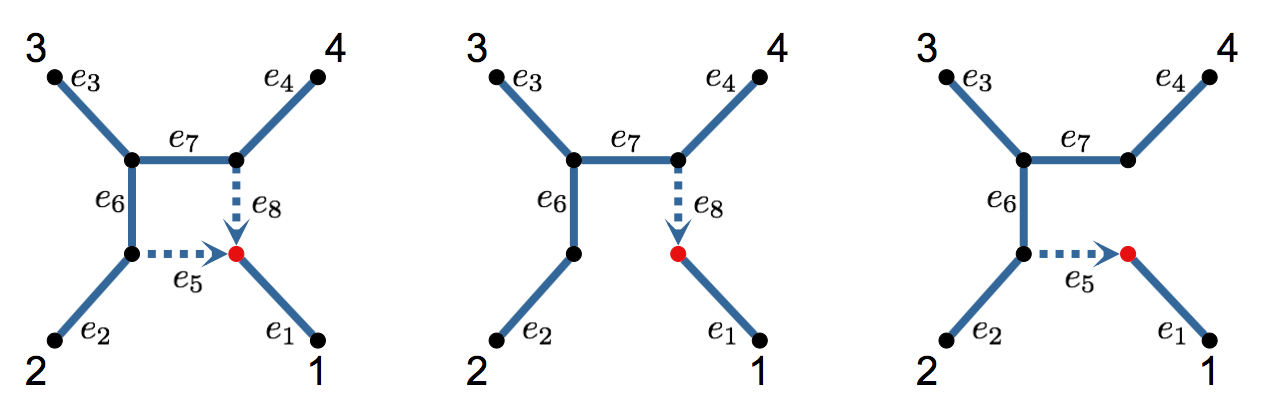}
		\end{center}
\end{figure}

The map $\phi_{N}$ sends
each variable in $R_n$ to a binomial
in $S_n$ where the two terms are determined
by $T_1$ and $T_2$. 
For what follows, let $L_m \subset [2n]$ be the set of edge indices of $T_m$.  For the variable
$q_{i_1,\ldots,i_n}$ the term that 
is associated to $T_m$ will be a monomial, with one indeterminate, $a^{(j)}_0$ or $a^{(j)}_1$, for each edge of the tree.  
In order to determine
the ``color" (0 or 1) of each edge indeterminate, we consider
$(i_1, \ldots, i_n)$ as a labeling of the leaves of $T_m$ by elements of 
$\mathbb{Z}_2$.
If we remove an edge $e_j$ of $T_m$, the resulting graph has two connected components which splits the leaves into two sets. 
Let $s^m_j(i_1,\ldots,i_n)$ be the group sum of the leaf labels on either side of the split induced by removing the edge $e_j$ from $T_m$.
(Note the sum of leaf labels is the same on
either side of the split.) 
The indeterminate associated to the  edge $e_j$ is then $a^{(j)}_{s^m_j(i_1, \ldots,i_n)}.$ 
Thus, we have the map

\[
\phi_N: R_n \to S_{n} \]
\[q_{i_1, \ldots, i_n} \mapsto 
\prod_{j \in L_1} a^{(j)}_{s^1_j(i_1,\ldots, i_n)} + 
\prod_{j \in L_2} a^{(j)}_{s^2_j(i_1,\ldots, i_n)}.
\]

\noindent Now the phylogenetic ideal $I_N$ associated to $N$ is the kernel of $\phi_N$:
$$I_{N} := ker(\phi_{N})= 
\{f \in R_n \ : \ \phi_N(f) = 0\}.$$

\begin{example}
\label{ex: edge coloring}
Let $N$ be the 6-leaf network pictured below.
Removing the reticulation edges of $N$ 
creates two trees, $T_1$ and $T_2$, with 
edge indices
$L_1 = [12]\setminus\{9\}$ and
$L_2 = [12]\setminus\{10\}$.
To determine the parameterization for the
coordinate $q_{111100}$, we color
the leaves by $(1,1,1,1,0,0)$. 
Here, we show vertices and edges colored by $1$ as magenta.

\begin{center}
    \includegraphics[width = 10cm]{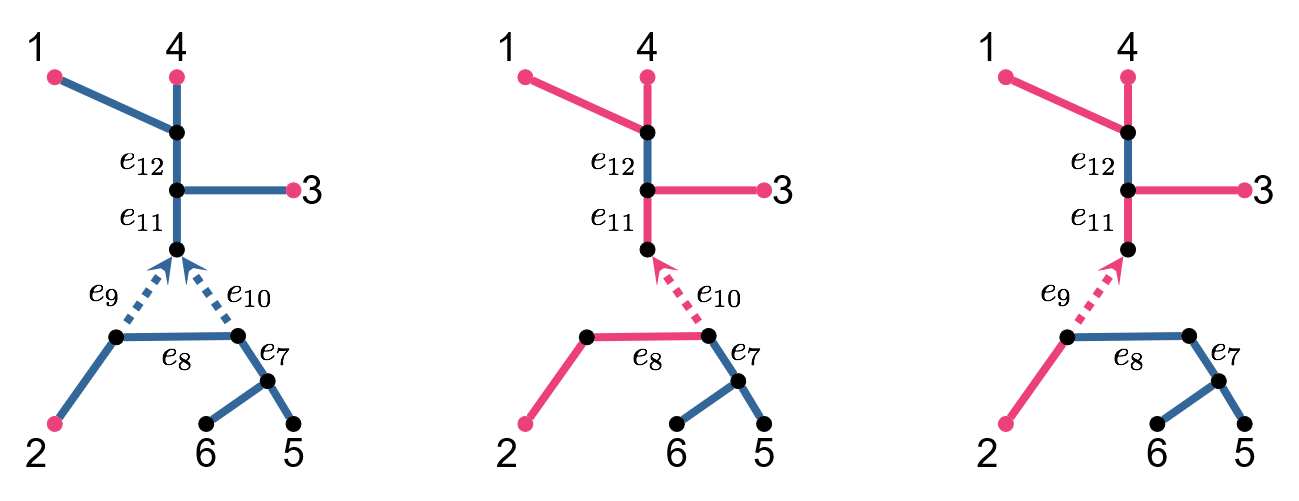}
\end{center}
\end{example}

To determine the color of an edge,
we sum the leaf colors on either side of the
split created by removing that edge. 
For example,
the edge $e_{11}$ in $T_1$ corresponds to the
split $134|256$. 
Since 
$$i_1 + i_3 + i_4 = 1 + 1 + 1 = 
i_2 + i_5 + i_6 = 1 + 0 + 0 = 1,$$
$e_{11}$ is colored by $1$. Thus,
the indeterminate for $e_{11}$ is $a^{(11)}_1$.
Then for the map $\phi_N$, we have

\begin{align*}
q_{111100} \mapsto \ &a^{(1)}_1a^{(2)}_1a^{(3)}_1a^{(4)}_1a^{(5)}_0a^{(6)}_0a^{(7)}_0a^{(8)}_1a^{(10)}_1a^{(11)}_1a^{(12)}_0 + \\
&a^{(1)}_1a^{(2)}_1a^{(3)}_1a^{(4)}_1a^{(5)}_0a^{(6)}_0a^{(7)}_0a^{(8)}_0a^{(9)}_1a^{(11)}_1a^{(12)}_0.
\end{align*}

\begin{exercise} 
Let $N$ be the quarnet with leaf label set $\{1,2,3,4\}$ from Exercise \ref{exer: LargeQuarnetSet}. 
Write out $R_n$, $S_n$, and $\phi_{N}$.  
Use Macaulay2 to compute $I_N$.
\end{exercise}

\begin{exercise}
Let $N$ be the sunlet network $N_k$. Show that the ring map in this section is the same ring map as the previous section if we replace the $b$ indeterminates with the appropriate $a$ indeterminates.
\end{exercise}

Notice that if we swap the leaf labels $1$ and $3$ in the 4-sunlet $N_4$, we obtain a different
4-sunlet network.
In the following challenge problem, we will see how changing the labeling of a network changes which polynomials are in the ideal.

\chproblem{How many labeled $4$-leaf sunlets are there? Compute $I_{N}$ for each of the 4-leaf sunlets.  How are the generating sets of each of these ideals related to $I_4$? }

For the sunlet graphs, we could factor all of the indeterminates corresponding to the leaf edges from the binomial but no other indeterminates.  In essence, we could write the image of every variable in $R_k$ as a monomial multiplied by a binomial. We can also do this for the $k$-cycle
networks, and as Example \ref{ex: edge coloring} 
shows, sometimes we can factor out many more
indeterminates. From that example, we could write
\begin{align*}
q_{111100} \mapsto \ &a^{(1)}_1a^{(2)}_1a^{(3)}_1a^{(4)}_1a^{(5)}_0a^{(6)}_0a^{(7)}_0a^{(11)}_1a^{(12)}_0
(a^{(8)}_1a^{(10)}_1 + a^{(9)}_1a^{(11)}). \\
\end{align*}

The following challenge problem is aimed to get at this general phenomenon for $k$-cycle networks.

\chproblem{Write out the map $\phi_N$ for several 4-leaf and 5-leaf $k$-cycle networks. For each graph, which edge indeterminates can you factor for every binomial in the map?  Can you describe the general pattern for $k$-cycle networks?}

In our explorations we have seen that different networks may induce different phylogenetic ideals.  The ideals of certain networks may contain the ideals of other networks with the same leaf set.  This suggests we might try to understand the relationship between ideal containment and the corresponding network structures.

\resproject{Draw all of the 5-leaf level-one networks. Which networks have the same ideal under the CFN model?  Which networks have ideals that are contained in one another? }

To explore the structure of the ideals you might use the Macaulay 2 command {\tt isSubset(J,I)} to determine if the ideal $J$ is contained in the ideal $I$.  Similarly, {\tt I == J} will tell you if two ideals are equal. 
In order to formalize the ideal containment structures you identify, it might be helpful to use a mathematical object called a partially-ordered set (poset). 
The definition of a poset as well as examples can be found  in Chapter 6 of \cite{keller2016applied}.

\bibliographystyle{plain}
\bibliography{references}

\begin{thebibliography}{10}

\bibitem{Agarwal2016Whoiswho}
Tushar Agarwal, Philippe Gambette, and David Morrison.
\newblock Who is who in phylogenetic networks: Articles, authors, and programs,
  2016.

\bibitem{Allman}
Elizabeth~S. Allman, Sonja Petrovi\'c, John~A. Rhodes, and Seth Sullivant.
\newblock Identifiability of 2-tree mixtures for group-based models.
\newblock {\em IEEE/ACM Trans. Comp. Biol. Bioinformatics}, 8(3):710--722,
  2011.

\bibitem{Allman2004Mathematical}
Elizabeth~S. Allman and John~A. Rhodes.
\newblock {\em Mathematical Models in Biology, an Introduction}.
\newblock Cambridge University Press, Cambridge, United Kingdom, 2004.

\bibitem{Allman2006}
Elizabeth~S. Allman and John~A. Rhodes.
\newblock The identifiability of tree topology for phylogenetic models,
  including covarion and mixture models.
\newblock {\em J. Comp. Biol.}, 13(5):1101--1113, 2006.

\bibitem{Gascuel2007}
Elizabeth~S. Allman and John~A. Rhodes.
\newblock {\em Reconstructing Evolution: New Mathematical and Computational
  Advances}, chapter~4.
\newblock Oxford University Press, UK, June 2007.

\bibitem{anaya2016determining}
Maria Anaya, Olga Anipchenko-Ulaj, Aisha Ashfaq, Joyce Chiu, Mahedi Kaiser,
  Max~Shoji Ohsawa, Megan Owen, Ella Pavlechko, Katherine~St. John, Shivam
  Suleria, Keith Thompson, and Corinne Yap.
\newblock On determining if tree-based networks contain fixed trees.
\newblock {\em Bulletin of Mathematical Biology}, 78(5):961--969, 2016.

\bibitem{Casanellas2006}
Marta Casanellas and Jes{\'u}s Fern{\'a}ndez-S{\'a}nchez.
\newblock Performance of a new invariants method on homogeneous and
  nonhomogeneous quartet trees.
\newblock {\em Molecular biology and evolution}, 24(1):288--293, 2006.

\bibitem{Cavender}
J.A. Cavender and Joseph Felsenstein.
\newblock Invariants of phylogenies in a simple case with discrete states.
\newblock {\em J. of Class.}, 4:57--71, 1987.

\bibitem{Chang1996}
J.T. Chang.
\newblock Full reconstruction of {M}arkov models on evolutionary trees:
  identifiability and consistency.
\newblock {\em Math. Biosci.}, 137(1):51--73, 1996.

\bibitem{chartrand2013first}
Gary Chartrand and Ping Zhang.
\newblock {\em A First Course in Graph Theory}.
\newblock Courier Corporation, 2013.

\bibitem{Cox2007}
David Cox, John Little, and Donal O'shea.
\newblock {\em Ideals, Varieties, and Algorithms}.
\newblock Undergraduate Texts in Mathematics. Springer Science+Business Media,
  third edition, 2007.

\bibitem{dress2012basic}
Andreas Dress, Katharina~T. Huber, Jacobus Koolen, Vincent Moulton, and Andreas
  Spillner.
\newblock {\em Basic Phylogenetic Combinatorics}.
\newblock Cambridge University Press, Cambridge, United Kingdom, 2012.

\bibitem{Eriksson2009}
Nicholas Eriksson.
\newblock Using invariants for phylogenetic tree construction.
\newblock In {\em Emerging applications of algebraic geometry}, pages 89--108.
  Springer, 2009.

\bibitem{Evans1993}
S.N. Evans and T.P. Speed.
\newblock Invariants of some probability models used in phylogenetic inference.
\newblock {\em Ann. Statist.}, 21(1):355--377, 1993.

\bibitem{Felsenstein2004}
Joseph Felsenstein.
\newblock {\em Inferring Phylogenies}.
\newblock Sinauer Associates, Inc., Sunderland, UK, 2004.

\bibitem{Francis2016}
Andrew Francis, Charles Semple, and Mike Steel.
\newblock New characterisations of tree-based networks and proximity measures.
\newblock {\em Advances in Applied Mathematics}, 93, Februrary 2018.

\bibitem{M2}
D.R. Grayson and M.E. Stillman.
\newblock Macaulay2, a software system for research in algebraic geoemetry.
\newblock Available at http://www.math.uiuc.edu/Macaulay2/, 2002.

\bibitem{Gross2017distinguishing}
Elizabeth Gross and Colby Long.
\newblock Distinguishing phylogenetic networks.
\newblock {\em SIAM J. Appl. Algebra Geometry}, 2(1):72--93, 2018.

\bibitem{grunewald2008quartet}
Stefan Gr{\"u}newald, Peter~J Humphries, and Charles Semple.
\newblock Quartet compatibility and the quartet graph.
\newblock {\em the electronic journal of combinatorics}, 15(1):103, 2008.

\bibitem{Hassett2007}
Brendan Hassett.
\newblock {\em Introduction to Algebraic Geometry}.
\newblock Cambridge University Press, New York, 2007.

\bibitem{Huber2018Quarnet}
Katharina~T. Huber, Vincent Moulton, Charles Semple, and Taoyang Wu.
\newblock Quarnet inference rules for level-1 networks.
\newblock {\em Bulletin of Mathematical Biology}, 80(8):2137--2153, August
  2018.

\bibitem{huson2010phylogenetic}
Daniel~H. Huson, Regula Rupp, and Celine Scornavacca.
\newblock {\em Phylogenetic Networks: Concepts, Algorithms and Applications}.
\newblock Cambridge University Press, Cambridge, United Kingdom, 2010.

\bibitem{VanIersel2014}
Leo~Van Iersel and Vincent Moulton.
\newblock Trinets encode tree-child and level-2 phylogenetic networks.
\newblock {\em J. Math Biol.}, 68(7):1707--1729, June 2014.

\bibitem{Jansson2006}
J.~Jansson, N.B. Nguyen, and W.K. Sung.
\newblock Algorithms for combining rooted triplets into a galled phylogenetic
  network.
\newblock {\em SIAM J. Comput.}, 35(5):1098--1121, 2006.

\bibitem{Jansson2006Inferring}
Jesper Jansson and Wing-Kin Sung.
\newblock Inferring a level-1 phylogenetic network from a dense set of rooted
  triples.
\newblock {\em Theoretical Computer Science}, 363(1):60--68, October 2006.

\bibitem{Keijsper2013}
Judith Keijsper and R.A. Pendavingh.
\newblock Reconstructing a phylogenetic level-1 network from quartets.
\newblock {\em Bulletin of Mathematical Biology}, 76(10):2517--2541, October
  2014.

\bibitem{keller2016applied}
Mitchel~T Keller and William~T Trotter.
\newblock {\em Applied Combinatorics}.
\newblock Mitchel T. Keller, William T. Trotter, 2016.

\bibitem{Long2015a}
Colby Long and Seth Sullivant.
\newblock Identifiability of 3-class {J}ukes--{C}antor mixtures.
\newblock {\em Advances in Applied Mathematics}, 64:89--110, 3 2015.

\bibitem{Maddison1997}
W.P. Maddison.
\newblock Gene trees in species trees.
\newblock {\em Syst. Biol.}, 46(523--536), 1997.

\bibitem{mirarab2014astral}
S.~Mirarab, R.~Reaz, M.S. Bayzid, T.~Zimmermann, M.S. Swenson, and T.~Warnow.
\newblock {ASTRAL}: genome-scale coalescent-based species tree estimation.
\newblock {\em Bioinformatics}, 30:i541--i548, 2014.

\bibitem{Nakhleh2011}
Luay Nakhleh.
\newblock {\em Problem Solving Handbook in Computational Biology and
  Bioinformatics}, chapter Evolutionary Phylogenetic Networks: Models and
  Issues, pages 125--158.
\newblock Springer Science+Business Media, LLC, 2011.

\bibitem{Pachter2005}
Lior Pachter and Bernd Sturmfels, editors.
\newblock {\em Algebraic Statistics for Computational Biology}, page 101.
\newblock Cambridge University Press, Cambridge, United Kingdom, 2005.

\bibitem{Pardi2015}
Fabio Pardi and Celine Scornavacca.
\newblock Reconstructible phylogenetic networks: Do not distinguish the
  indistinguishable.
\newblock {\em PLos Comput Biol.}, 11(4), April 2015.

\bibitem{Rusinko2012}
Joseph~P. Ruskino and Brian Hipp.
\newblock Invariant based quartet puzzling.
\newblock {\em Algorithms Mol Biol.}, 7(35), 2012.

\bibitem{Semple2016}
Charles Semple.
\newblock Phylogenetic networks with every embedded phylogenetic tree a base
  tree.
\newblock {\em Bulletin of Mathematical Biology}, 78(1):132--137, 2016.

\bibitem{Semple2003}
Charles Semple and Mike Steel.
\newblock {\em Phylogenetics}.
\newblock Oxford University Press, Oxford, 2003.

\bibitem{snir2012quartet}
S.~Snir and S.~Rao.
\newblock Quartet {M}ax{C}ut: a fast algorithm for amalgamating quartet trees.
\newblock {\em Molecular Phylogenetics and Evolution}, 62(1):1--8, January
  2012.

\bibitem{Solis2016}
Claudia Sol\'is-Lemus and C\'ecile An\'e.
\newblock Inferring phylogenetic networks with maximum pseudolikelihood under
  incomplete lineage sorting.
\newblock {\em PLOS Genetics}, 2016.

\bibitem{Steel2016Phylogeny}
Mike Steel.
\newblock {\em Phylogeny: Discrete and Random Processes in Evolution}.
\newblock CBMS-NSF Regional Conference Series in Applied Mathematics. SIAM,
  2016.

\bibitem{sturmfels1996grobner}
Bernd Sturmfels.
\newblock {\em Gr{\"o}bner bases and Convex Polytopes}, volume~8.
\newblock American Mathematical Soc., 1996.

\bibitem{Sturmfels2005}
Bernd Sturmfels and Seth Sullivant.
\newblock Toric ideals of phylogenetic invariants.
\newblock {\em J. Comp. Biol.}, 12(2):204--228, 2005.

\bibitem{Sturmfels2006}
Bernd Sturmfels and Seth Sullivant.
\newblock Combinatorial secant varieties.
\newblock {\em Quarterly Journal of Pure and Applied Mathematics}, 2:285--309,
  2006.

\bibitem{Sullivant2018algebraic}
Seth Sullivant.
\newblock {\em Algebraic Statistics}.
\newblock Graduate Studies in Mathematics. American Mathematical Society, 2018.

\bibitem{syvanen1994}
M.~Syvanen.
\newblock Horizontal gene transfer: evidence and possible consequences.
\newblock {\em Annu. Rev. Genet.}, 28:237--261, 1994.

\end{thebibliography}

\end{document}